\documentstyle [prl,aps]{revtex}

\begin{document}

%\twocolumn[\hsize\textwidth\columnwidth\hsize\csname @twocolumnfalse\endcsname
\title{Induced gravity with complex metric field}

\author{ W.F. Kao\thanks{email:wfgore@cc.nctu.edu.tw}}
\address{Institute of Physics, Chiao Tung
University, Hsinchu, Taiwan}

\maketitle

% Abstract 
\begin{abstract}
The possible existence of a complex metric tensor field is studied.
We show that an effective scalar field is induced by an overall phase component of the complex metric tensor.
The corresponding gauge field is shown to be a tachyon.
Possible implications of this scalar field to the no hair theorem in a spherically symmetric space is also analyzed. 
We also study its impact on the evolution of the early universe.
\end{abstract} \vskip .2in

PACS numbers: 04.20.-q; 04.20.Cv; 04.70.Bw; 98.80.Cq;  

\section{Introduction}

It is known that the metric tensor field associated with a general Riemannian space \cite{book} is real by construction. 
Recently, the complex extension of the Einstein gravity for the non-commutative gravity \cite{cpg} has attracted lots of attention.
The complex metric may or may not be realized by our physical nature.
It is, however, interesting by itself to investigate the possible existence of a complex counterpart for the real metric field in curved space \cite{cpg}.
In this paper, we will take a different approach and extract a simple phase factor out of this extended complex metric field.
We will hence study the possible impact of this phase field.

One can simply generalize the real metric field to a complex metric field by assuming
${\bf g}_{\mu \nu}= {\rm g}_{\mu \nu} +i \gamma_{\mu \nu}=g_{\mu \nu}\exp
[ i \theta_{\mu \nu} ]$ with both
${\rm g}_{\mu \nu}$ and $\gamma_{\mu \nu}$ indicating real symmetric tensor fields.
The invariant length can be defined solely by the modulus, 
$g_{\mu \nu}$ of the complex metric field 
${\bf g}_{\mu \nu}$, i.e. $ds^2 \equiv g_{\mu \nu}dx^\mu dx^\nu$.
In addition, the inverse metric ${\bf g}^{\mu \nu}$ can be defined as the inverse of the matrix 
${\bf g}$ where $({\bf g})_{\mu \nu} \equiv {\bf g}_{\mu \nu}$.
In another word, we define 
${\bf g}^{\mu \nu}$ by ${\bf g}^{\mu \nu} {\bf g}_{\nu \alpha}
= \delta^\mu_\alpha$ with $\delta^\mu_\alpha$ the Kroneker delta.

If we define the spin connection as the original Christoffel symbol with
$g_{\mu \nu}$ replaced by ${\bf g}_{\mu \nu}$, 
one expects that the effective linearized real Lagrangian density for the complex metric tensor 
field associated with the Einstein-Hilbert action $-\sqrt{{\bf g}} R$ should read
\begin{equation}
 {\cal L} ={1 \over 4} {\bf h}^{\mu \nu} \partial^2 {\bf h}_{\mu \nu}
\label{cL} \end{equation}
when one choose the Einstein gauge Re $[\partial_\mu {\bf h}^{\mu
\nu}]=0$.
Here we have assumed the expansion ${\bf g}_{\mu \nu} = \eta_{\mu \nu} +{\bf
h}_{\mu \nu}$ against the Minckowski background.
Note further that one can only perform gauge (coordinate) transformation with four parameters in $4$-dimensional real space.
Therefore, one can only take care of four of the graviton fields when a gauge is chosen.
In order to treat the complex metric theory similar to the real metric theory, one would have to extend the real spacetime to a complex spacetime.
This is however not our approach.

In order to work on a real-valued Lagrangian density, one would have to take the real part of the complex Einstein-Hilbert Lagrangian as the effective Lagrangian. 
The real part consists, however, of an original graviton field and another ghost field with a negative kinetic energy term according to Eq. (\ref{cL}).
Hence it appears that a straightforward generalization of the complex-extension of the graviton field will not bring in much physical impact.

On the other hand, it is comparably easy to extract information about a real scalar field associated with the common phase factor $\phi$ of the phase factors by assuming 
$\theta_{\mu \nu} =\phi$ for all $\mu$, $\nu$.
In another word, we propose to work on a simple generalization of the metric tensor by assuming
${\bf g}_{\mu \nu}= {\rm g}_{\mu \nu} +i \gamma_{\mu \nu}=g_{\mu \nu}\exp [ i \phi ]$ for all $\mu$, $\nu$.
To be more specifically, we will assume that
$\theta_{\mu \nu} =\phi$ for all $\mu, \nu$.
The resulting effective Lagrangian can be shown to give
\begin{equation}
- \sqrt{{\bf g}} {\bf R} = - \sqrt{g}\exp [ i\phi ] \;
  [R  +{3 \over 2} \partial_\mu \phi \partial^\mu \phi ] .
\label{actionphi}
\end{equation}
One can take the effective Lagrangian as the real part of Eq. (\ref{actionphi}) by averaging it with its complex conjugate. 
The final result reads
\begin{equation}
{\cal L}_{\rm eff} \equiv {1 \over 2} [ - \sqrt{{\bf g}} {\bf R} +{\rm c.c.}]=
- \sqrt{g} \cos \phi  \;
[R  +{3 \over 2} \partial_\mu \phi \partial^\mu \phi ] .
\label{action-eff}
\end{equation}
Note that the effective Lagrangian is invariant under a discrete transformation 
$\phi \to \phi +2n\pi$ for all $n \in \bf{Z}$.
In addition, we will assume that the scalar field is bounded in a cycle around some minimum $\phi_0$ by, for example,
$|\phi-\phi_0| < \pi/2$.
In fact, we need to ensure that the gravitational constant related to $\cos \phi$ remains finite for all era when this effective theory is in charge.
One also notes that the coefficient of the inverse of the gravitational constant $1/G \sim  \cos \phi \to 1- \phi^2/2$ if $\phi$ is close to zero.
The theory in this limit has been the focus of recent activities \cite{ab}.

A more detailed investigation shows that the overall common phase field $\phi$ associated with the complex metric field resembles the Weyl's measuring field \cite{scale,kaocbh,kaoinflation}.
In fact, this $\phi$ field behaves similar to the imaginary extension of the Weyl's original idea. 
Therefore, one can introduce the associated gauge transformation by 
$\phi_\theta = \phi +\theta$ such that ${\bf g}_{\mu \nu}^\theta =
{\bf g}_{\mu \nu} \exp [i \theta ]$. 
There then requires the existence of a gauge field associated with this phase transformation such that 
$A_\mu^\theta = A_\mu + \partial_\mu
\theta$. 
One must also replace all derivative in the real action with covariant derivative, namely, $\partial_\mu g_{\alpha \beta} \to \nabla _\mu g_{\alpha \beta}
\equiv (\partial_\mu -i A_\mu) g_{\alpha \beta}$.
The real Hilbert-Einstein action from Eq. (\ref{actionphi}) can then be shown to give an effective quadratic interaction term of the following form
$3 A_\mu A^\mu/2 $.
This introduces a negative mass-square term for the gauge field $A_\mu$.
Hence the associated gauge field is a tachyon.
This indicates that the gauge extension from this method does not appear to be helpful unless a tachyon is in need.
But the effective theory without gauge extension still provides a natural source of the scalar field one is looking for. 

For the fermionic part, one would have
\begin{equation}
{\cal L}_{\psi} \equiv {1 \over 2} [  \sqrt{{\bf g}}
(-i \bar{\Psi} \gamma^\mu{\bf D}_\mu \Psi)
+{\rm c.c.}]=
\sqrt{g} \left \{  \exp [ i {3 \over 2} \phi] \;
 [ -i \bar{\psi} \gamma^\mu{D}_\mu \psi  +{9 \over 4} \partial_\mu \phi  J^\mu ] 
+ c.c. \right \}
\label{action-psi}
\end{equation}
by writing $\Psi = \exp[i3\phi/2] \psi$ for any spin $1/2$ spinor field $\psi$.
Here $D_\mu \psi \equiv  (\partial_\mu +{1\over 2}\sigma^{bc}
    e^\nu_b  D_\mu e_{\nu c} ) \psi$,
with $e^\mu_a$ denoting the vierbein such that $e^\mu_a e^\nu_b \eta^{ab}
= g^{\mu\nu}$ \cite{book}. 
Here $a,b,c$ denote flat  space indices.
Moreover, $J^\mu \equiv i \bar{\psi} \gamma^\mu \psi$ is the charged current for the fermion field $\psi$.

%One can further show that the total effective-Lagrangian becomes
%\begin{equation}
%{\cal L}_{\rm tot} \sim
%- \sqrt{g} [1- {\phi^2 \over 2}] \;\;
%[ R  +{3 \over 2} \partial_\mu \phi \partial^\mu \phi ]
%+\sqrt{g} \left \{  [ 1-{9 \over 8} \phi^2]
%\left[ -i \bar{\psi} \gamma^\mu{D}_\mu \psi + {1 \over 2} D_\mu J^\mu
%\right ]
%+ {9 \over 16}  \partial_\mu  \phi^2 J^\mu \right \}
%\label{action-effpsi}
%\end{equation}
%under the small field expansion where $\phi \ll 1$.

One remarks here that by writing $\Psi = \exp[i3\phi/2] \psi$, one simply splits the fermionic field into the product of a phase factor and a physical fermionic field.
We do this because there is no clear guideline for us to do the separation.
The only difference will be on the coefficient of the last term in Eq. (\ref{action-psi}) if we choose, for example, 
$\Psi = \exp[ik\phi] \psi$ for $k \ne 3/2$.
In fact the coefficient will be $k+3/4$ instead of $9/4$.
In addition, the choice $k =3/2$ we made make the extracted phase factor behave similar to a complex generalization of the Weyl theory.
To be more specifically, 
$\Psi_{\Lambda, \theta} = \Psi \exp[3(\Lambda+i\theta)/2]$ under a combined complex Weyl transformation
${\bf g}_{\mu \nu}^{\Lambda, \theta} = {\bf g}_{\mu \nu} \exp[(2\Lambda+i\theta)]$ \cite{scale,kaocbh,kaoinflation}.
.

Note further that the last term is a derivative coupling that has been used in the study of $\pi$-$N$ scattering \cite{ryder} if a chiral current
$J_\mu^5$ is present. 
This kind of chiral interaction will also be induced when a $\gamma^5$ coupling is present in the fermionic Lagrangian ${\cal L}_\psi$.
Note that the $\gamma^5$ matrix is defined to be a pseudo scalar by 
$\gamma^5 \equiv \epsilon_{\mu \nu \alpha \beta} \gamma^\mu \gamma^\nu \gamma^\alpha \gamma^\beta /4!$.
Therefore, the complex extension of the $\gamma^5$ matrix does not contain any phase field ($\phi$) contribution.
Hence, one can directly add chiral term in above fermionic equations.
Note again that he coefficient of the derivative coupling term will be different if we choose, for example, 
$\Psi = \exp[ik\phi/2] \psi$.
The similarity shown above indicates that there is something important hidden in the complex metric extension of the Einstein theory.

One can put it in another way by saying that the scalar filed can be recombined to act as a phase component of the complex graviton field. 
Nonetheless, we expect that the resulting scalar field can be an alternative for the physical evolution of our early universe.
In order to learn more about the impact of this model, we will study a few simple models both in a spherically symmetric space and in a homogeneous and isotropic Friedmann-Robertson-Walker (FRW) space \cite{book,data} . 

One knows that many symmetries may not be realized by nature in the low energy regime \cite{scale,kaocbh}. 
Hence, we will introduce an effective symmetry breaking potential $V(\phi)$ to the system.
In short, one will be working with the following model:
\begin{equation}
{\cal L}_{\rm eff}=
- \sqrt{g} \cos \phi  \;
[R  +{3 \over 2} \partial_\mu \phi \partial^\mu \phi ] -V.
\label{action-effV}
\end{equation}
One can show that the equation of motion becomes
\begin{eqnarray}
\cos \phi \; G_{\mu \nu} + (g_{\mu \nu} D^2 -D_\mu \partial_\nu ) \; \cos \phi
&=& {3 \over 2} \partial_\mu  \phi \partial_\nu  \phi \cos \phi- 
 { 1 \over 2} g_{\mu \nu} [ {3 \over 2} (\partial \phi )^2\cos \phi +V]
\\
\sin \phi \, [R+{3 \over 2} (\partial \phi )^2] + 3 D^\mu ( \cos \phi \partial_\mu \phi)
&=& V_\phi
\end{eqnarray}
Here $G_{\mu \nu} \equiv g_{\mu \nu} R/2 -R_{\mu \nu}$ denotes the Einstein tensor.
One can also derive the following equation from eliminating the $R$-term in the trace of the $g_{\mu \nu}$-equation and the $\phi$ equation. 
The result is quite similar to the conventional constraint equation in the scale invariant theory \cite{kaocbh}:
\begin{equation}
D^2 \phi = {1 \over 3} [ V_\phi \cos \phi  + 2V \sin \phi ].
\end{equation}
Note that a symmetry breaking potential of the form $V=V(\cos \phi)$ will be introduced in the course of proving the no-hair theorem for a spherically symmetric space.
Hence we will denote $V'={\partial V(\cos \phi) / \partial \cos \phi}$ and 
$V_\phi={\partial V(\cos \phi) / \partial \phi}$ respectively from now on throughout the rest of this paper.
In addition, we will assume that the scalar field is bounded in a cycle around a minimum $\phi_0$ by, for example,
$|\phi-\phi_0| < \pi/2$.
This is to ensure that the gravitational constant related to $\cos \phi$ remains finite for all era when this effective theory is in charge.
Following Ref. \cite{kaocbh}, we will assume a class of symmetry breaking potential as scaling potential if the potential
$V$ satisfies the inequality $(\cos \phi - \cos \phi_0) [ V' \cos \phi -2V] >0$ for some constant $\phi_0$, and for all $\phi$.
One can show that the no-hair theorem holds for this scalar field $\phi$ under the effect of this sort of scaling potential.
Indeed, the proof is quite similar to the proof shown in Ref. \cite{kaocbh}.
The spherically symmetric metric is given by
$ds^2 \equiv g_{\mu \nu} dx^\mu dx^\nu = -\exp [2A(r)] dt^2 + \exp [2B(r)] dr^2 +r^2 d \Omega$ in this paper.
To be more specifically, one can show that
\begin{equation}
3\int_{r_H}^\infty dr r^2 \exp [A-B] \sin \phi \phi'^2 
+ \int_{r_H}^\infty dr r^2 \exp [A+B] \sin \phi [ \cos \phi - \cos \phi_0] [ V' \cos \phi -2V] =0.
\end{equation}
Note that both integrands are positive (negative) provided that $\sin \phi >0$ ($<0$).
Hence, both integrands have to be vanishing throughout all space exterior to the event horizon at $r=r_H$ where $\exp[A-B]|_{r_H} =0$.
Therefore, one shows that the no-hair theorem holds for all scaling potentials provided that $\sin \phi$ is bounded from approaching zero.
Note that one can show that the symmetry breaking potential 
$V_4 =(\cos \phi - \cos \phi_0)^2$ and $\cos ^2 \phi  \ln [ \cos ^2 \phi /\cos^2 \phi_0] - (\cos^2 \phi - \cos^2 \phi_0)$ are all scaling potentials.
In these models one can easily constrain the scalar field from touching zeros of $\sin \phi$.
One also notes that in the limit $\phi \sim 0$ and $\phi_0 \sim 0$, the potential
$V_4 =(\cos \phi - \cos \phi_0)^2$ reduces to the standard symmetry breaking potential
$(\phi^2 - \phi_0^2)^2/4$.
Hence the symmetry breaking potential $V_4$ introduced here is nothing more than a simple generalization of the standard $\phi^4$ SSB potential.

On the other hand, one has
\begin{eqnarray}
H_0^2 \cos \phi_0 &\sim& V(\phi_0)/6 \\
\sin \phi_0 \, [2H_0^2] +  \cos \phi_0 \; 3H_0 \dot{\phi}_0/2 
&=& -V_\phi/6 
\end{eqnarray}
in the FRW space under slow rollover approximation such that
$H_0 \gg |\dot{\phi}_0|$ and $H_0 |\dot{\phi}_0 | \gg |\ddot{\phi}_0|$.
Here the FRW metric is given by $ds^2= -dt^2 + a^2(t) [ dr^2 /(1-k/r^2) +r^2 d\Omega]$ with $H \equiv \dot{a}/a$ the Hubble parameter.
In addition, $H=H_0$ denotes a de Sitter space.
Assuming that $\sin \phi_0 \sim 0$, one can show that $H_0^2 \sim V(\phi_0)/6$ and
$\dot{\phi} \sim -V_\phi /9H_0$. 
This is similar to the ordinary gravity system studied in the literature.

In summary, one has shown that a class of complex metric field induces a phase field that behaves very much the way conventional scalar field.
The generalization of this work to a more general form is still under investigation.
The model studied here is a generalization of the Weyl's conformal theory.
The associated gauge field for the phase symmetry is, however, a tachyon.
Therefore, the gauge symmetry corresponding to the phase generalization of the real metric field may not be gauged unless tachyon is in need.
We extract, however, the effective real action from the bosonic sector and study its implication to the no-hair theorem and inflationary universe.
It is shown that, as long as the scalar field one introduced is bounded to a single cycle of the corresponding cosine functional, this effective theory behaves similar to a conventional induced gravity model.
Further applications to other fields of interest will be presented in a sequel to this paper.

\vspace{0.2in}

\section{Acknowledgments}
%{\bf \large Acknowledgments}
This work was supported in part by NSC under the contract number
NSC88-2112-M009-001.

\end{document}